\documentclass[conference]{IEEEtran}

\newenvironment{LimitationBox}{\begin{tcolorbox}[colback=yellow!10,colframe=red!80!black,title=Limitation]}{\end{tcolorbox}}
\newenvironment{ContributionBox}{\begin{tcolorbox}[colback=green!10,colframe=green!80!black,title=Contribution]}{\end{tcolorbox}}
\IEEEoverridecommandlockouts

\usepackage{todonotes}
\usepackage{cite}
\usepackage{amsmath,amssymb,amsfonts}
\usepackage{algorithmic}
\usepackage{graphicx}
\usepackage{textcomp}
\usepackage{xcolor}
\usepackage{enumitem}
\usepackage{subcaption}
\usepackage{standalone}
\usepackage{tikz}
\usepackage{float}
\usepackage{censor}
\StopCensoring

\usepackage[switch]{lineno}
\usepackage{booktabs,tabularx}
\usepackage{placeins}
\newcommand{\panel}[1]{\includegraphics[width=\linewidth]{#1}}
\newcommand{\Tsub}[1]{T_{\mathrm{#1}}}
\newcommand{\tsub}[1]{t_{\mathrm{#1}}}

\usepackage{listings}
\definecolor{codegray}{HTML}{6F6A7A}
\definecolor{codebg}{HTML}{F8F7FA}
\definecolor{codeink}{HTML}{3A245C}
\definecolor{coderule}{HTML}{D8D2E2}
\lstdefinestyle{modelfile}{
  basicstyle=\ttfamily\scriptsize,
  backgroundcolor=\color{codebg},
  commentstyle=\color{codegray}\itshape,
  keywordstyle=\color{codeink}\bfseries,
  stringstyle=\color{codegray},
  frame=single, framerule=0.3pt, rulecolor=\color{coderule},
  breaklines=true, breakatwhitespace=true, postbreak=\mbox{\ \ },
  columns=fullflexible, keepspaces=true, showstringspaces=false,
  aboveskip=6pt, belowskip=6pt,
  title=\ttfamily\footnotesize\color{codegray}\lstname}
\lstdefinestyle{platformxml}{style=modelfile, language=XML,
  markfirstintag=true, tagstyle=\color{codeink}\bfseries,
  usekeywordsintag=false}
\lstdefinestyle{shimconf}{style=modelfile,
  comment=[l]{\#},
  morekeywords={mpi.device_path,gdr.latency,gdr.bandwidth,
    model.enabled,gpu.flops_rate,gpu.flops_per_thread,
    h2d.latency,h2d.bandwidth,d2h.latency,d2h.bandwidth,
    d2d.latency,d2d.bandwidth,peer.path,staging.share,
    exchange.overhead,op.bw,memcpy.bw,allreduce.call_overhead}}
\usepackage{booktabs}
\usepackage{colortbl}

\newcommand{\rnum}[1]{\raisebox{0.15ex}{\textcircled{\scriptsize #1}}}

\usetikzlibrary{positioning,fit,arrows.meta,calc,backgrounds}

\usepackage[most]{tcolorbox}

\tcbset{width=\columnwidth,boxrule=0pt,colback=red,arc=0pt,auto outer arc,left=0pt,right=0pt,boxsep=5pt}

\definecolor{customcolor}{HTML}{7C2E76}

\newtcolorbox[auto counter]{ObservationBox}{
    borderline west={3pt}{0pt}{DeepSkyBlue4},
    colback=customcolor!30!white}

\usepackage[hidelinks]{hyperref}
\usepackage{xurl}  
\def\BibTeX{{\rm B\kern-.05em{\sc i\kern-.025em b}\kern-.08em
    T\kern-.1667em\lower.7ex\hbox{E}\kern-.125emX}}

\def\BibTeX{{\rm B\kern-.05em{\sc i\kern-.025em b}\kern-.08em
    T\kern-.1667em\lower.7ex\hbox{E}\kern-.125emX}}

\begin{document}

\title{Orrery: Modeling and Simulation of GPU-direct MPI Communication with SimGrid}

\newif\ifshowauthors
\showauthorstrue

\ifshowauthors
\author{
\IEEEauthorblockN{Ahmad Moh'd Saleh A. Belbeisi}
\IEEEauthorblockA{
\textit{Technical University of Munich}\\
Munich, Germany\\
ahmad.belbeisi@tum.de
}
\and
\IEEEauthorblockN{Amir Raoofy}
\IEEEauthorblockA{
\textit{Leibniz Supercomputing Centre}\\
Munich, Germany\\
amir.raoofy@lrz.de
}
\and
\IEEEauthorblockN{Josef Weidendorfer}
\IEEEauthorblockA{
\textit{Dresden University of Technology}\\
Dresden, Germany\\
Josef.Weidendorfer@tu-dresden.de
}

\and[\hfill\mbox{}\par\vspace{1em}\mbox{}\hfill]
\IEEEauthorblockN{Majid Salimi Beni}
\IEEEauthorblockA{
\textit{Vienna University of Technology}\\
Vienna, Austria\\
majid.salimibeni@tuwien.ac.at
}
}
\fi

\maketitle
\begin{abstract}

Simulation of HPC applications is an essential tool for analyzing system design trade-offs and enabling effective hardware--software co-design.
As modern supercomputers increasingly rely on GPU acceleration, MPI implementations have evolved to support GPU-aware communication, enabling direct GPU data transfers.
Modeling these mechanisms is therefore critical for simulating HPC applications on current systems.
In this paper, we introduce Orrery, an interposition layer built on SimGrid's SMPI to model GPU-aware MPI communication using existing collective implementations and flow-based simulation.
Orrery intercepts GPU memory allocations, kernel launches, data transfers, and MPI operations, and models their execution in SimGrid according to target system specifications, enabling simulation without requiring physical GPUs.
We validate Orrery using CUDA-aware MPI microbenchmarks and a Jacobi solver on up to 32 nodes of the \censor{MUSICA} supercomputer and demonstrate that it reproduces key GPU-aware MPI mechanisms.

\end{abstract}


\section{Introduction}
\label{sec:intro}
GPUs are widely used in modern HPC systems, and, e.g., in the latest TOP500 ranking (June 2026)~\cite{top500}, four of the top five supercomputers are equipped with GPUs. Consequently, MPI implementations have evolved to support GPU-aware communication, enabling direct interaction between GPU memory and distributed communication mechanisms. 
As GPU-accelerated systems become increasingly prevalent, accurate performance modeling and simulation of these architectures are essential for analyzing, understanding, and optimizing HPC applications, as well as for guiding the design of future systems.
Simulators leveraging novel techniques and tools play a critical role in evaluating the performance and scalability of applications and algorithms:

Simulation enables evaluation of system configurations and application designs, allowing system
architects and application developers to rapidly evaluate algorithm and hardware scalability and compare alternative approaches before costly deployments on large-scale systems.

Simulating modern GPU-accelerated HPC applications requires capturing and modeling kernel executions, data paths and data movements, as well as the support for GPU-aware communication mechanisms, often hidden behind the MPI interface.
In particular, modeling GPU-aware MPI, empowered by peer-to-peer GPU communications~\cite{6587715} that is often used in modern GPU-accelerated systems,
becomes increasingly important for simulating many scientific applications. SimGrid~\cite{4488918,casanova2025lowering} provides a strong foundation for this goal through its execution-driven model, detailed network abstraction, and SMPI support for simulating MPI applications; however, SMPI has been designed for CPU-based MPI execution and lacks native support for representing GPU memory management, GPU kernel execution, and GPU-aware MPI communication. 

This work addresses this gap by extending SimGrid with GPU-aware communication capabilities, in particular targeting a CUDA-aware platform with GPUDirect RDMA~\cite{nvidia_gpudirect_rdma} transports. We introduce \textit{Orrery}\footnote{Orrery and the calibration artifacts are publicly available under the MIT
license at \url{https://github.com/ahmadbelb/orrery}.}, an interposition layer that enables the online simulation of CUDA-aware MPI applications. Without modifying SimGrid, \textit{Orrery} intercepts and models GPU memory management operations like allocation and data transfer calls, kernel launches, and GPU-aware MPI communications, allowing for leveraging existing SimGrid and SMPI simulation engines for GPU-enabled HPC applications.

The main contributions of this work are as follows: 
\begin{enumerate}[label=\textbf{C\arabic*:}, ref=C\arabic*,
                  leftmargin=*, align=left, labelsep=0.5em, itemsep=0.3ex]

\item We present Orrery, an interposition layer built on the SMPI, and its first implementation, the first GPU-aware extension in SimGrid ecosystem. 
that enables online simulation of CUDA-aware MPI applications. 
\item We introduce two execution modes enabling applications to be simulated regardless of whether GPU devices are physically present on the simulation platform (Twin mode) or not (Skeleton mode).
\item We provide an initial validation of Orrery on up to 32 nodes (128 GPUs) of the \censor{MUSICA} supercomputer, using CUDA-enabled MPI micro-benchmarks and a GPU-enabled Jacobi solver application. 
\end{enumerate}

\section{Background and Related Work}
\label{sec:background}
\noindent \textbf{SimGrid:} SimGrid is an open-source framework for simulating distributed systems through analytical models of computation, communication, and I/O over user-defined virtual platforms. Its SMPI module enables execution-driven simulation of MPI applications by redirecting MPI operations to SimGrid's implementation, where each rank is represented as a lightweight simulation actor. Communication operations are simulated using SimGrid's flow-level network models and SMPI’s implementations of the collective algorithms of MPI libraries, such as the ones implemented in Open\,MPI \cite{smpi}, while computations are represented by either native kernel execution and scaling their measured durations to the simulated platform, or by replacing them with virtual delays (advancing per rank virtual clocks). This approach enables lightweight and scalable 
simulation of MPI applications on a single machine without requiring access to the target system~\cite{smpi}.

\noindent \textbf{GPU-aware MPI}:
Unlike traditional MPI, GPU-aware MPI accepts device pointers as communication buffers, allowing data to remain on the GPU unless staging through host memory is enabled~\cite{6587715}. 

GPU-aware MPI implementations typically determine the memory type of communication buffers and select an appropriate data path, including peer-to-peer transfers over PCIe or NVLink within a node, GPUDirect RDMA across nodes~\cite{nvidia_gpudirect_rdma}, or staged transfers through pinned host memory when direct GPU communication is unavailable.

\noindent \textbf{Related Work}:
Prior work has explored simulation approaches for large-scale HPC systems, primarily focusing on CPU-based communication and network behavior.
Trace-driven simulators capture, parse, and replay application executions
under different system configurations; for example, LogGOPSim~\cite{hoefler2010loggopsim} and PHANTOM~\cite{10.1145/1837853.1693493} provide MPI-oriented communication models, while ATLAHS~\cite{10.1145/3712285.3759838} enables evaluation of distributed workloads through a common representation and multiple simulation backends.
However, trace-based approaches capture only a specific execution instance, requiring new traces for different configurations and limiting scalability studies~\cite{10.1145/3712285.3759838,cornebize2022simulation}, while trace collection and storage can become costly at large scale~\cite{10.1145/3458817.3476151,10.1145/3712285.3759838}. 
Execution-driven simulators, 
such as SimGrid/SMPI~\cite{smpi} and xSim~\cite{5999835}, address these limitations by executing applications against a modeled platform, while SST/macro~\cite{janssen2010simulator} and BigSim~\cite{1303013} combine both approaches.
With the rise of GPU-accelerated systems, simulators such as ASTRA-sim~\cite{astra1,astra2,astra3} and SimAI~\cite{wang2025simai} have extended simulation toward distributed AI workloads, through specialized GPU communication tools such as NCCL~\cite{hu2026demystifyingnccl}. None of these approaches support GPU-aware MPI: execution-driven simulators lack device-memory awareness, while AI-oriented simulators primarily target NCCL-based communications rather than MPI.

\section{Simulating GPU-aware MPI with Orrery}
\label{sec:orrery}
\begin{figure}[b]
    \centering
    \includegraphics[scale=1]{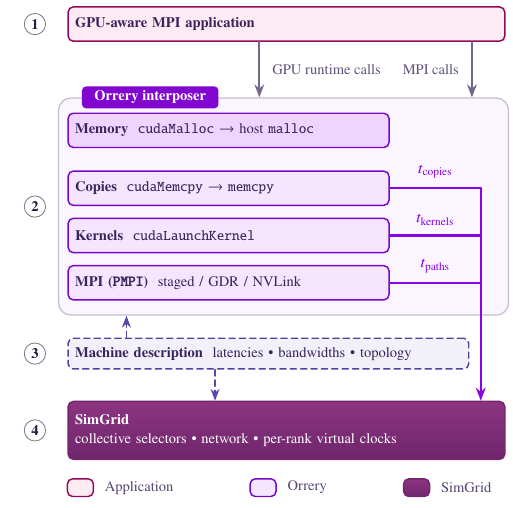}
  \caption{Orrery overview, depicting how it intercepts GPU runtime and MPI calls to leverage SimGrid's simulation engine.}
\label{fig:interposer}
\end{figure}
Figure~\ref{fig:interposer} presents the overall architectural design of Orrery.
Orrery acts as an interposition layer that interfaces with SMPI to support GPU-aware MPI applications without modifying the underlying SimGrid framework.
An application leveraging a GPU-aware MPI~\rnum{1} runs unmodified, while Orrery~\rnum{2} intercepts relevant GPU (CUDA) runtime and MPI calls. GPU memory allocations, data transfers, kernel executions, and communication operations are captured and modeled as a target machine description representing the bandwidth and latencies of operations and the data paths~\rnum{3}. 
This design allows using the existing SimGrid simulation cores: 
The resulting execution costs are charged to each rank's virtual clock within SimGrid~\rnum{4}, which continues to provide the underlying process scheduling and network simulation.

\subsection{Orrery Interposition}

SMPI's standard build process targets host-based MPI applications and cannot compile or link device code. Orrery extends the build process by using the vendor toolchain (\texttt{nvcc} in our prototype) for device code compilation and final linking while retaining SMPI's MPI interface. The interposition layer and cost model (Sec.~\ref{sec:cost-model}) are compiled with the application into a shared object loaded by \texttt{smpirun}, making interception a link-time property rather than a runtime preload mechanism. To intercept CUDA runtime operations, Orrery provides CUDA wrappers bound at the application binary interface (ABI): link-time symbol binding redirects calls such as \texttt{cudaMalloc} to Orrery without parsing or transforming the source. MPI interception leverages the standard MPI Profiling Interface (PMPI) mechanism~\cite{mpi-standard}: Orrery provides weak MPI wrappers, while \texttt{PMPI\_*} symbols are resolved by SMPI. Therefore, both call paths follow application $\rightarrow$ Orrery wrappers $\rightarrow$ SMPI/SimGrid, enabling CUDA operations and MPI communication to be captured and modeled in a unified simulation environment.

\subsection{Modeling GPU-direct P2P, Collectives, and Kernels}
\label{sec:cost-model}
Each intercepted operation is assigned a duration derived from the machine description and added to the rank's virtual clock. Data movement follows the standard latency--bandwidth model established by Hockney and LogGP~\cite{DBLP:journals/pc/Hockney94,DBLP:journals/simpra/HoeflerSL09}.
The \emph{charging structure} defines which costs a GPU-aware call incurs, which rank is charged, and whether each cost is applied per call or per message.
A call's cost is composed of recurring terms. The central term is the \emph{device leg}: the first or last segment between GPU memory and the network interface. Charged once per device buffer on the owning rank, its cost follows the selected transport:

\begin{equation}
\label{eq:leg}
\tsub{leg}(n) =
\begin{cases}
  \tsub{d2h}(n) + \tsub{h2d}(n) & \text{Host staging} \\[2pt]
  \alpha_{\mathrm{gdr}}         & \text{GPUDirect RDMA} \\[2pt]
  \tsub{peer}(n)                & \text{NVLink / PCIe peer} \\[2pt]
  0                             & \text{Host-resident buffer},
\end{cases}
\end{equation}
where $\tsub{d2h}$, $\tsub{h2d}$, and $\tsub{peer}$ follow piecewise-linear latency--bandwidth models, while $\alpha_{\mathrm{gdr}}$ represents the per-message overhead of direct NIC access to device memory.

Calls combine device legs with network time. A point-to-point transfer pays one leg per endpoint. Collectives (e.g., broadcast, allgather) pay one leg per device buffer involved—one for broadcast and two for allgather—plus a fixed one per-call overhead. Operator-based collectives (allreduce, reduce) are also covered~\cite{mpi-standard}; in the case of Open\,MPI's CUDA component~\cite{ompi_coll_cuda}, operands are staged to host memory, reduced on the host, and copied back, replacing device legs with host bounce and reduction costs:
\begin{equation}
\label{eq:comm}
\begin{aligned}
\Tsub{p2p}  &= \tsub{leg}(n_{\mathrm{s}}) + \Tsub{net} + \tsub{leg}(n_{\mathrm{r}}), \\
\Tsub{coll} &= \tsub{leg}(n_{\mathrm{s}}) + \tsub{leg}(n_{\mathrm{r}})
               + \tsub{call} + \Tsub{net}, \\
\Tsub{coll}^{\mathrm{op}}
            &= {\underbrace{2\,\tsub{d2h} + \tsub{h2d}}_{\text{host bounce}}}
               + \tsub{reduce} + \tsub{call} + \Tsub{net}.
\end{aligned}
\end{equation}
Here $n_{\mathrm{s}}$ and $n_{\mathrm{r}}$ denote send and receive payload sizes. $\tsub{call}$ represents the per-call overhead, while $\tsub{reduce} = (P-1)\,n/(P\,\beta_{\mathrm{op}})$ models the reduce-scatter arithmetic floor, where $P$ is the communicator size and $\beta_{\mathrm{op}}$ is the host reduction rate~\cite{DBLP:journals/ijhpca/ThakurRG05}. $\Tsub{net}$ is provided by SimGrid: Orrery accounts for GPU-side costs and invokes \texttt{PMPI\_*}, while SimGrid models the communication algorithm on the declared topology.

Where needed, a kernel launch is modeled as a fixed overhead plus device work:
\begin{equation}
\label{eq:kernel}
\begin{aligned}
\Tsub{kernel}  &= \tsub{launch} + \frac{Nf}{R}, &
\Tsub{compute} &= \sum_{\ell} \Tsub{kernel}^{(\ell)}
                  + \sum_{s} \tsub{sync}^{(s)} ,
\end{aligned}
\end{equation}
\noindent
where $N = G \times B$ is the total thread count, $f$ the per-thread work, and $R$ the achieved device rate. $\tsub{launch}$ and $\tsub{sync}$ are measured launch and synchronization overheads, with $\ell$ and $s$ indexing kernel launches and synchronization calls.

\subsection{Machine Description and Model Calibration}
\label{sec:calibration}

The machine description provides the parameters of the cost model
(Eqs.~\ref{eq:leg}--\ref{eq:kernel}). Since achieved performance often
differs from theoretical peaks, inflating the error~\cite{DBLP:conf/sc/McDonaldWMSSPDC25}, each parameter is
calibrated using a dedicated micro-benchmark on the reference cluster.
Table~\ref{tab:calib} summarizes the probes used in these micro-benchmarks and their target parameters.

\begin{table}[H]
  \centering
  \setlength{\abovecaptionskip}{4pt}%
  \setlength{\belowcaptionskip}{2pt}%
  \caption{Calibration probes and collected model parameters.}
  \label{tab:calib}
  \footnotesize
  \setlength{\tabcolsep}{4pt}
  \renewcommand{\arraystretch}{1.0}
  \setlength{\aboverulesep}{0.2ex}
  \setlength{\belowrulesep}{0.3ex}
  \begin{tabular}{@{}llr@{}}
    \toprule
    \textbf{Probe} & \textbf{Yields} & \textbf{Symbol} \\
    \midrule
    GPU-GPU intra-node    & link latency, bandwidth        & $\tsub{peer}$ \\
    GPU-GPU inter-node   & wire latency, bandwidth        & $\Tsub{net}$ \\
    MPI API overhead     & per-call library cost         & $\alpha_{\mathrm{gdr}}$ \\
    Host--device copy        & copy latency, banded bandwidth & $\tsub{d2h}$, $\tsub{h2d}$ \\
    Empty kernel launch  & launch, sync overheads         & $\tsub{launch}$, $\tsub{sync}$ \\
    Kernel compute            & per-kernel compute rate        & $R$ \\
    \texttt{MPI\_Op} time & host operation (e.g., reduction) time      & $\beta_{\mathrm{op}}$ \\
    \bottomrule
  \end{tabular}
\end{table}

\subsection{Execution Modes}
Orrery provides two execution modes, following SMPI's treatment of computation phases. In both modes, intercepted operations are evaluated by the cost model and their durations are added to the rank's virtual clock. SimGrid handles scheduling and contention: Orrery determines the operation duration, while SimGrid determines its execution.

The modes differ only in how wrappers handle operations before charging:

\noindent\textbf{Skeleton mode} does not require GPUs: memory allocations use an internal registry, and kernels and data transfers are modeled with cost models. However, the device buffers contain no output, and no kernel is launched. This mode uses all Equations~\ref{eq:leg}-\ref{eq:kernel} for modeling.

\noindent\textbf{Twin mode} requires a GPU for GPU kernel execution (does not use Equation~\ref{eq:kernel}), and forwards each call to the CUDA runtime
and charges the measured duration. This is done by leveraging
\texttt{-Bsymbolic} to resolve the name to the wrapper; the real entry
point comes from \texttt{dlsym(RTLD\_NEXT)}.

\section{Experimental Results}
\label{sec:validation}

\subsection{Experimental Setup}
\label{sec:setup}

Reference measurements are collected on 32 nodes of the \censor{MUSICA} supercomputer, each equipped with four fully connected NVIDIA H100 GPUs via NVLink\,4.0 and connected by 200\,Gbit/s InfiniBand. The system runs Open\,MPI 4.1.6 with UCX 1.15, CUDA support, and GPUDirect RDMA.

Simulations use skeleton mode on a workstation with an 4-core Intel Xeon W-2125 (4.0\,GHz), 32\,GB memory, and a 5\,GB Quadro P2000 GPU, running SimGrid 4.1.1.

\begin{figure}[!bp]
\centering
\begin{subfigure}[t]{0.49\linewidth}
    \panel{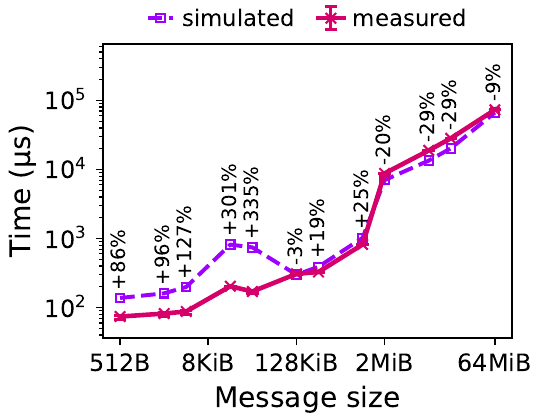}
    \caption{Allreduce.}
    \label{fig:gdr-sweep-allreduce}
\end{subfigure}\hfill%
\begin{subfigure}[t]{0.49\linewidth}
    \panel{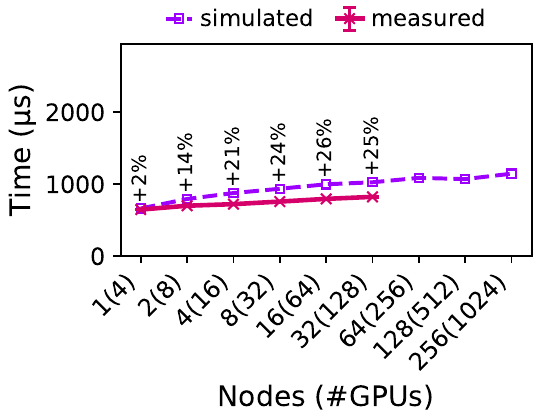}
    \caption{Allreduce scaling.}
    \label{fig:gdr-scaling-allreduce}
\end{subfigure}\\[0.4ex]
\begin{subfigure}[t]{0.49\linewidth}
    \panel{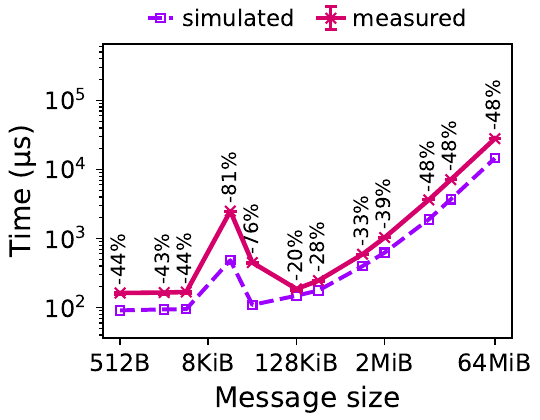}
    \caption{Bcast.}
    \label{fig:gdr-sweep-bcast}
\end{subfigure}\hfill%
\begin{subfigure}[t]{0.49\linewidth}
    \panel{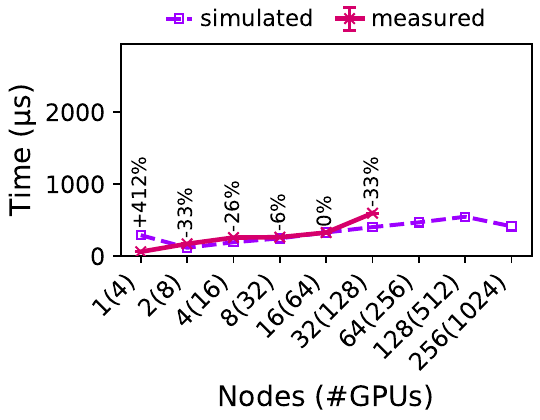}
    \caption{Bcast scaling.}
    \label{fig:gdr-scaling-bcast}
\end{subfigure}\\[0.4ex]
\begin{subfigure}[t]{0.49\linewidth}
    \panel{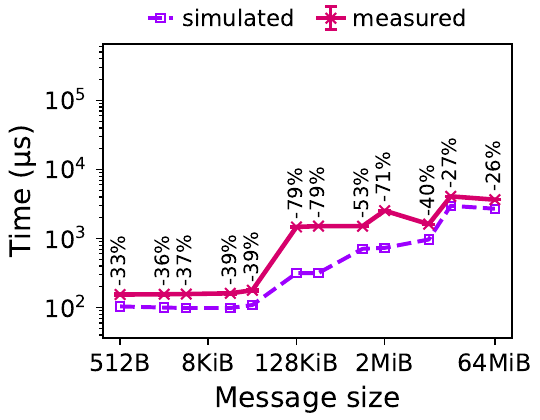}
    \caption{Allgather.}
    \label{fig:gdr-sweep-allgather}
\end{subfigure}\hfill%
\begin{subfigure}[t]{0.49\linewidth}
    \panel{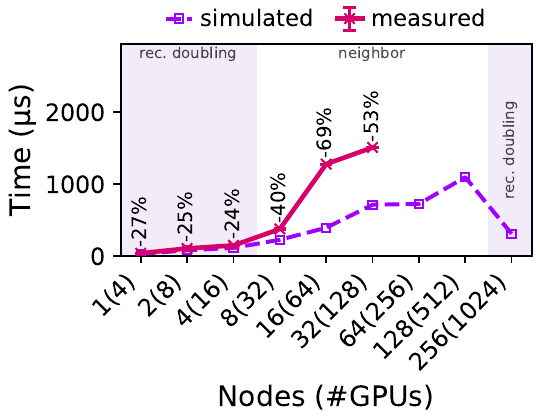}
    \caption{Allgather scaling.}
    \label{fig:gdr-scaling-allgather}
\end{subfigure}
\caption{Micro-benchmarks, measured versus simulated: The left column (a, c, and e) shows results on 32 nodes (128 GPUs); the right column (b, d, and f) shows strong scaling with a fixed message size of $2^{20}$\,B, from 1 node to 32 nodes. 
Relative simulation error is annotated next to each point, and the shaded regions in (f) highlight the collective algorithm selected by Open MPI at each scale.}
\label{fig:musica-six-gdr}
\end{figure}

\begin{figure}[!tp]
\centering
\begin{subfigure}[t]{0.49\linewidth}
    \panel{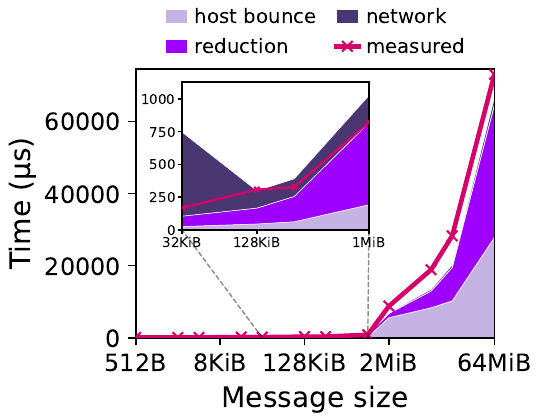}
    \caption{Allreduce.}
    \label{fig:gdr-attribution-allreduce}
\end{subfigure}\hfill%
\begin{subfigure}[t]{0.49\linewidth}
    \panel{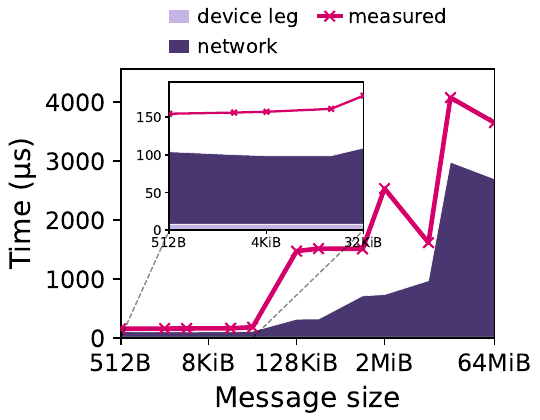}
    \caption{Allgather.}
    \label{fig:gdr-attribution-allgather}
\end{subfigure}
\caption{Simulated collective micro-benchmarks at 32 nodes (128 GPUs), decomposed into the costs the interposer charges.}
\label{fig:gdr-attribution}
\end{figure}

\subsection{Micro-benchmark Evaluations}
Figure~\ref{fig:musica-six-gdr} compares skeleton-mode simulation with micro-benchmark measurements. The left column varies message size at 128 GPUs: Orrery qualitatively reproduces GPU-direct MPI communication behavior on different collective operations and message sizes, with an average error of 60\%. The right column scales a fixed 1\,MiB message size up to 1024 GPUs for the simulator: The simulator can qualitatively reproduce scalability, and we can also see clearly the effect of Open\,MPI's algorithm selection for Allgather (Fig. \ref{fig:gdr-scaling-allgather}) where performance improves at 1024 GPUs with choosing the recursive doubling algorithm.

Figure~\ref{fig:gdr-attribution} provides a more detailed view
with simulated time
broken down into cost components. 

For Allreduce (~\ref{fig:gdr-attribution-allreduce}), host bounce and reduction time grow with the message size because the reduction operations are performed on the host, and data must be copied back and forth between host and device. 
Allgather (~\ref{fig:gdr-attribution-allgather}) and Bcast (not shown here) have no reduction and no such bottleneck.

\subsection{Jacobi Solver}
\begin{figure}[!tp]
\centering
\begin{subfigure}[t]{0.49\linewidth}
    \panel{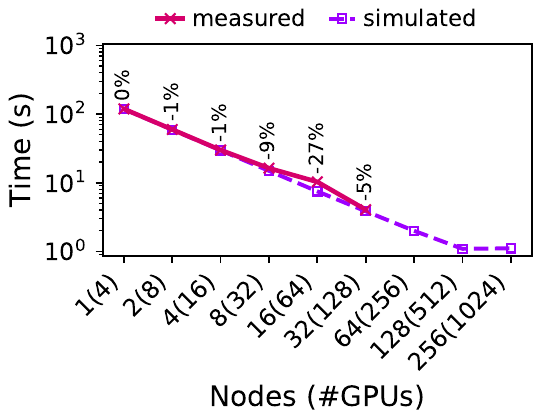}
    \caption{Strong scaling.}
    \label{fig:gdr-jacobi}
\end{subfigure}\hfill%
\begin{subfigure}[t]{0.49\linewidth}
    \panel{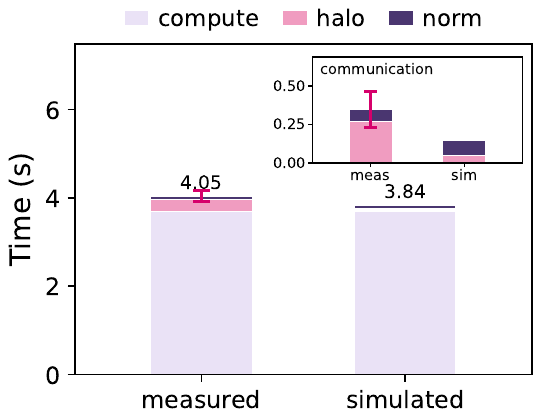}
    \caption{Decomposed runtimes.}
    \label{fig:gdr-app2}
\end{subfigure}
\caption{Jacobi with GPUDirect RDMA. (a) Measured versus simulated,
measured to 32 nodes (128 GPUs), simulated beyond. (b) The 128-GPU run split
into the solver's phases; the measured bar's range is the spread across
ranks. Inset: the two communication phases, magnified.}
\label{fig:gdr-applications}
\end{figure}

Figure~\ref{fig:gdr-applications} shows a Jacobi solver~\cite{HertenOGKHAM26}, simulated in skeleton mode with every
kernel launch modeled using Eq.~\ref{eq:kernel}. On different number of nodes (Fig.~\ref{fig:gdr-jacobi}), the relative error of the simulation is at most 27\%, and we can see how increasing the number of nodes in the simulator to 256 improves the performance. To further analyze the application, Fig.~\ref{fig:gdr-app2} decomposes the 128-GPU run into solver phases. The simulated and measured total run times differ by about 5\%, as compute dominates and its simulation error is only 0.1\%. The inset highlights the communication phases: 0.35\,s measured versus 0.14\,s simulated, with halo exchange undercharged ($-81\%$) and norm calculation overcharged ($+16\%$).

\section{Discussion}
\label{sec:discussion-extensions}
The errors in the simulation arise in part from calibration mismatches: the calibration relies on isolated two-GPU device-buffer measurements, whereas real applications involve different buffer types and communication patterns. In addition, collective operations introduce concurrent transfers that are not captured by the current calibration benchmarks and probes (listed in Table~\ref{tab:calib}). Part of the error is also structural: the cost model is built from simple latency–bandwidth terms in the style of Hockney \cite{DBLP:journals/pc/Hockney94} and LogGP \cite{DBLP:journals/simpra/HoeflerSL09}, which abstract away effects such as protocol switches, memory registration, and contention among concurrent transfers, so a residual error may remain even with well-tuned parameters. Improving accuracy requires independent measurements for host-buffer transfers and concurrent communication scenarios, as well as accounting for network and memory contention.

Two limitations remain in our current prototype: First, skeleton node relies on a single measured kernel rate, which is sufficient for memory-bound workloads but cannot distinguish computation and memory effects or generalize across kernels and architectures. Second, Orrery currently models operations sequentially, so it cannot capture computation-communication overlap in asynchronous applications. 

\section{Conclusion and Future Directions}
\label{sec:conclusion}
We introduced Orrery, an interposition layer that extends SimGrid's SMPI to GPU-aware MPI applications, modifying neither the simulator nor the application code.
It captures GPU runtime calls (e.g., CUDA calls) and charges each from a calibrated cost model, exposing the device-side cost of a GPU-aware MPI implementation to the simulation framework.

Several extensions can broaden Orrery's applicability. Starting with validation on full applications. The interception approach can be extended beyond CUDA to other runtimes such as HIP and SYCL, which would allow validating the cost model on GPUs from other vendors, while additional and automated calibration~\cite{DBLP:conf/sc/McDonaldWMSSPDC25} can improve model accuracy. Twin mode enables validation across real platforms, and the machine description allows exploration of future GPU systems, topologies, and extending previous MPI algorithm selection studies to be GPU-aware~\cite{10740853}. Such modeling can also evaluate emerging approaches, such as in-network computing, as in~\cite{belbeisi2026tofino}.

As the GPU communication stack moves further off the host, 
we envision Orrery setting the foundations for modeling GPU-aware communication based on GPUDirect RDMA in the well-established community-driven SimGrid simulator to better evaluate machines, libraries, and algorithms, including those that do not yet exist.

\section*{Acknowledgements}
The computational results presented in this work have been achieved using the Austrian Scientific Computing (ASC) infrastructure (cluster: MUSICA) under project ID 201169. Development and initial validation were carried out on systems in the BEAST (Bavarian Energy Architecture \& Software Testbed) test environment at the Leibniz Supercomputing Centre.

\bibliographystyle{IEEEtran}
\bibliography{main}

\end{document}